\documentclass{aastex631}
\usepackage{amsmath}

\begin{document}

\title{Passing Stars as an Important Driver of Paleoclimate and the Solar System's Orbital Evolution}

\author[0000-0001-5272-5888]{Nathan A. Kaib}
\affiliation{Planetary Science Institute, 1700 E. Fort Lowell, Suite 106, Tucson, AZ 85719, USA}
\affiliation{Homer L. Dodge Department of Physics \& Astronomy, University of Oklahoma, 440 W. Brooks St, Norman, OK 73019, USA}

\author[0000-0001-8974-0758]{Sean N. Raymond}
\affiliation{Laboratoire d'Astrophysique de Bordeaux, CNRS and Universit\'e de Bordeaux, All\'ee Geoffroy St. Hilaire, 33165 Pessac, France}



\begin{abstract}

Reconstructions of the paleoclimate indicate that ancient climatic fluctuations on Earth are often correlated with variations in its orbital elements. However, the chaos inherent in the solar system's orbital evolution prevents numerical simulations from confidently predicting Earth's past orbital evolution beyond 50--100 Myrs. Gravitational interactions among the Sun's planets and asteroids are believed to set this limiting time horizon, but most prior works approximate the solar system as an isolated system and neglect our surrounding Galaxy. Here we present simulations that include the Sun's nearby stellar population, and we find that close-passing field stars alter our entire planetary system's orbital evolution via their gravitational perturbations on the giant planets. This shortens the timespan over which Earth's orbital evolution can be definitively known by a further $\sim$10\%. In particular, in simulations that include an exceptionally close passage of the Sun-like star HD 7977 2.8 Myrs ago, new sequences of Earth's orbital evolution become possible in epochs before $\sim$50 Myrs ago, which includes the Paleocene-Eocene Thermal Maximum. Thus, simulations predicting Earth's past orbital evolution before $\sim$50 Myrs ago must consider the additional uncertainty from passing stars, which can open new regimes of past orbital evolution not seen in previous modeling efforts. 

\end{abstract}



\section{Introduction} \label{sec:intro}

Over the past 3--4 decades, computational dynamical simulations have enabled direct modeling of the orbital evolution of the solar system over Myr--Gyr timescales \citep{lask88, quinn91, wishol91}. This work has demonstrated that the orbital evolution of the Sun's inner planets are chaotic, with Lyapunov timescales of $\sim$5 Myrs, and this chaos is thought to be driven by the proximity of multiple secular planetary frequencies as well as asteroid-asteroid encounters \citep{susswis88, lask89, lask90, lask11}. Because we do not know the modern solar system's properties to infinite precision, chaotic divergence prevents us from knowing the planets' detailed orbital evolution to arbitrarily long times, and it is only characterized in a statistical sense beyond $\sim$100-Myr timescales, as a broad range of behavior is possible \citep{lask94}. In fact, even the long-term stability of the solar system is not guaranteed over the Sun's lifetime, as simulations indicate a $\sim$1\% chance that Mercury will be lost via collision with the Sun or Venus \citep{lask08}. 

Consequently, Earth's past or future orbital evolution can only be confidently predicted inside a time horizon much shorter than Earth's age, as small uncertainties in current planetary orbits eventually lead to dramatically divergent behavior. The time horizon set by the internal chaos among the Sun's eight planets is $\sim$70 Myrs, but additional strong chaos resulting from encounters between large asteroids shortens the horizon by another $\sim$10 Myrs \citep{lask11}. However, inside this time horizon, backwards integration of the Sun's planets has been used to predict the detailed past orbital evolution of the Earth \citep{lask04}. Because variations in the Earth's paleoclimate are known to correlate with Earth's eccentricity fluctuations \citep{mil41, hays76}, this has allowed us to reconstruct the chronology of sedimentation records and assess the orbital context of notable climate events \citep{zeeb19}. 

Simulations of the long-term orbital evolution of the Sun's planets have nearly always modeled the solar system as an isolated system. For many purposes, this is a very good approximation, but the solar system is of course part of the Milky Way galaxy. Consequently, it occasionally suffers close encounters with other field stars \citep{opik32, oort50}, and solar neighborhood kinematic studies predict an average of $\sim$20 stellar passages within 1 pc of the Sun each Myr \citep{bail18}. Because the solar system cross-section scales with the square of heliocentric distance, the large majority of these encounters will be distant and inconsequential to the planets' dynamics, but this is not guaranteed. In fact, there is a $\sim$0.5\% chance that a field star passage will trigger the loss of one or more planets over the next 5 Gyrs \citep{brownrein22, ray23}, and such passages may actually guarantee the disruption of the planets' orbits many Gyrs after the Sun becomes a white dwarf \citep{zink20}. Yet, encounters need not trigger an instability for them to have dynamical consequences for the planets. For instance, it has been suggested that $\sim$1/3 of Neptune's modern eccentricity has been generated through past stellar encounters \citep{brun93}, but many of the long-term dynamical effects of stellar passages remain unknown. Here we use numerical simulations to characterize how stellar passages resulting from the Sun's local galactic environment affect the chaotic evolution of our planets. Our work and its results are described in the subsequent sections. 

\section{Methods}

We use a modified version of the MERCURY hybrid integrator to perform the simulations presented here \citep{cham99}. This modified version allows for the inclusion of an arbitrary number of stellar mass bodies \citep{kaib18} and also features a simplified central force modification to induce general relativistic orbital precession \citep[Equation 30 of][]{sahatre92}. Unless specified otherwise, the simulations here are integrated for 150 Myrs with a step size of 1.5 days. Orbital elements of the planets and Pluto are set to the heliocentric osculating January 1, 2000 elements specified within the JPL Horizons system. The same is done for asteroids Ceres, Vesta, Pallas, Iris, and Bamberga when asteroids are included (which is noted when this is the case).  To generate unique sets of initial conditions, the mean anomaly of each body is altered by a random amount that generates an orbital drift between $\pm2$ cm, which is obviously much smaller than the actual uncertainty of the bodies' positions and orbits. 

Some of our simulations include populations of passing field stars. In these cases, stellar passages are initiated at random positions 1 pc away from the Sun with an isotropic velocity distribution \citep{hen72}. Stellar masses are assigned with the present-day stellar mass function \citep{reid02}, and the passage frequency and velocity distributions of each stellar spectral class are specified based on characterizations of solar neighborhood kinematics \citep{gar01, rick08}. Our stellar passage routine generates 18 stellar encounters within 1 pc per Myr, which compares favorably with recent observational estimates \citep{bail18}. However, only passages with impact parameters below 0.1 pc are actually integrated, as more distant passages require extraordinarily high masses or low velocities to generate significant perturbations (see results). Passing stars are integrated until their heliocentric distance again exceeds 1 pc, at which point the body is removed from the simulation. 

In simulations that ``neglect'' stars, we actually still introduce the same sets of stellar passages from our non-isolated runs. The only difference is that the stellar masses are scaled down by a factor of 1000. This is done to try to keep the numerical algorithm sequences between batches of runs as similar as possible, since it is known that the dynamical behavior of a simulated solar system can display variance with the numerical algorithm's properties \citep{zeeb15}.

Earth's Moon and the Sun's oblateness are excluded from our simulations, as well as any tidal or rigid body effects. Our simulations are not designed to generate the most accurate history of Earth's orbital evolution, and far more sophisticated models exist \citep{lask04, lask11b, zeeb19}. Instead, they are designed to highlight and characterize the important effects that stellar passages have on the orbital evolution of the planets and compare these effects to those of asteroids as well as the internal chaos that planet-planet interactions drive. 

\subsection{Secular Mode Analysis}

To decompose a planet's orbital eccentricity evolution into its eigenfrequencies and solve for their phases and amplitudes, we use the Modified Fourier Transform available at 
\href{https://www.boulder.swri.edu/~davidn/fmft/fmft.html}{https://www.boulder.swri.edu/$\sim$davidn/fmft/fmft.html} \citep{sidnes96, lask99}.  For the frequencies, amplitudes, and phases measured in Figure \ref{fig:example}, orbital elements are recorded every 100 years for 3.2768 Myrs. In all other cases discussed here, we record orbital elements every 100 years for 6.5536 Myrs.

\section{The Effects of Stellar Passages} \label{sec:style}

\begin{figure}
\centering
\includegraphics[width=.8\linewidth]{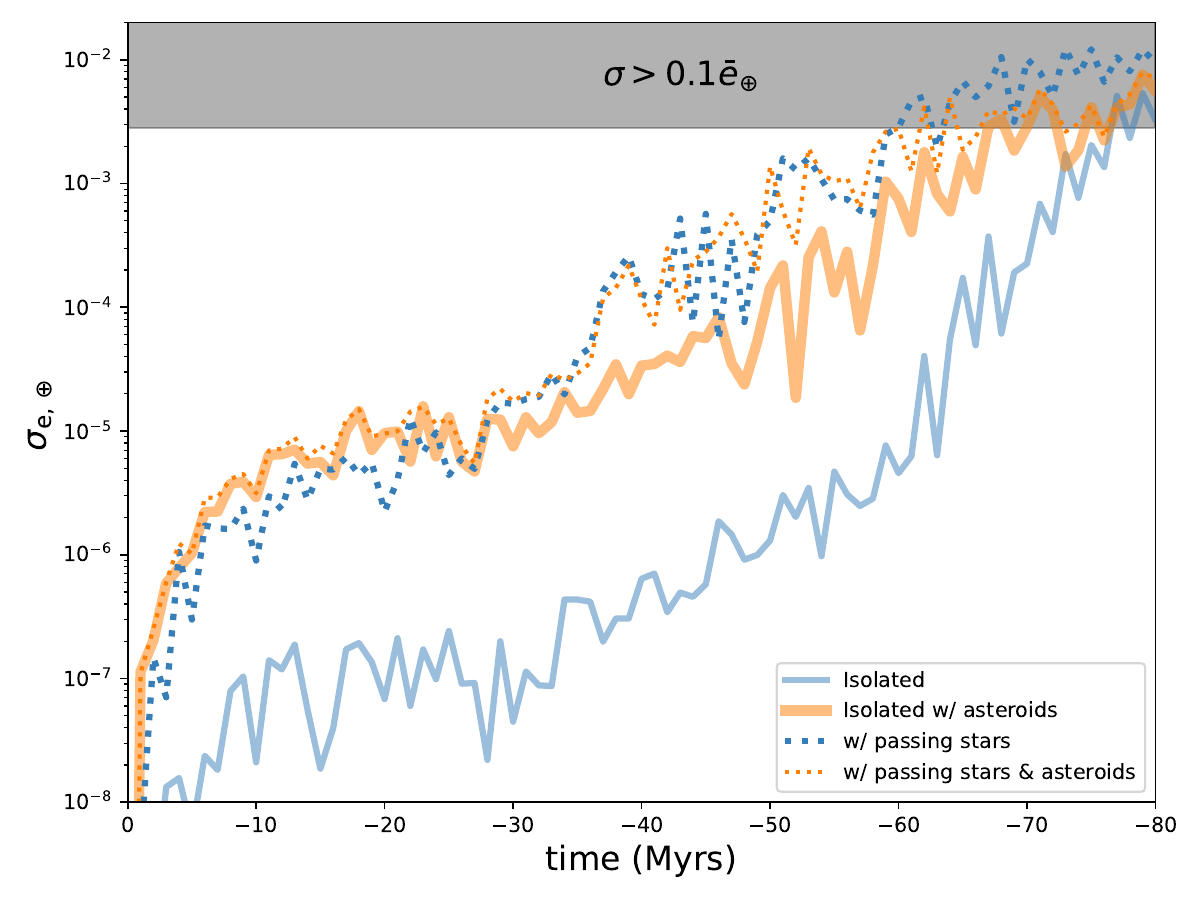}
\caption{The standard deviation of Earth's orbital eccentricity is plotted against time for 100 back-integrations of solar system clones. Four batches of solar system simulations are run. Two only consider the Sun, its eight planets, and Pluto ({\it blue lines}), and the other two ({\it orange lines}) also include 5 large asteroids: Ceres, Vesta, Pallas, Iris, and Bamberga. Half of our batches also perturb the solar system with passing field stars ({\it dashed lines}), while the others do not include this perturbation ({\it solid lines}).}
\label{fig:eccsigma}
\end{figure}

Given that even asteroid-asteroid interactions are known to accelerate the chaotic diffusion of Earth's orbit, it may be that perturbations from other stars can also accelerate the divergence of orbital evolution predictions. In Figure \ref{fig:eccsigma}, we demonstrate that stellar passages do indeed significantly accelerate Earth's orbital evolution divergence. Here we back-integrate 100 realizations (or clones) of the modern solar system for 150 Myrs and measure the standard deviation of Earth's orbital eccentricity among these systems over time. \citep[Eccentricity divergences beyond 10\% of Earth's mean eccentricity, or 0.0028, have been previously used to mark the time beyond which an orbital solution is unreliable][]{zeeb17}. If we only consider the eight planets and Pluto, we find that this standard deviation reaches 10\% of Earth's mean orbital eccentricity after 77 Myrs of back-integration. While our simulations neglect some effects considered in other works such as the Sun's oblateness, tidal dissipation, and gravity from Earth's moon, this result is roughly consistent with previous results \citep{lask11}. When we next repeat the integrations including the asteroids Ceres, Vesta, Pallas, Iris, and Bamberga, we find that this time horizon shortens by 10 Myrs down to 67 Myrs, and this decrease in divergence time is in approximate agreement with prior results \citep{lask11}. However, when we then repeat these back-integrations in the presence of a population of passing stars based on the solar neighborhood's properties \citep{rick08, garc01}, we find that the time horizon again shortens by another 5--7 Myrs (62 Myrs in the case with stellar passages and the asteroids, and 60 Myrs in the case with stellar passages but without asteroids). Because we see a significantly shorter time horizon whether we include asteroids or not, we conclude that perturbations from stellar passages degrade our ability to predict Earth's past or future orbital evolution even more strongly than asteroid perturbations. 

\begin{figure}
\centering
\includegraphics[width=.7\linewidth]{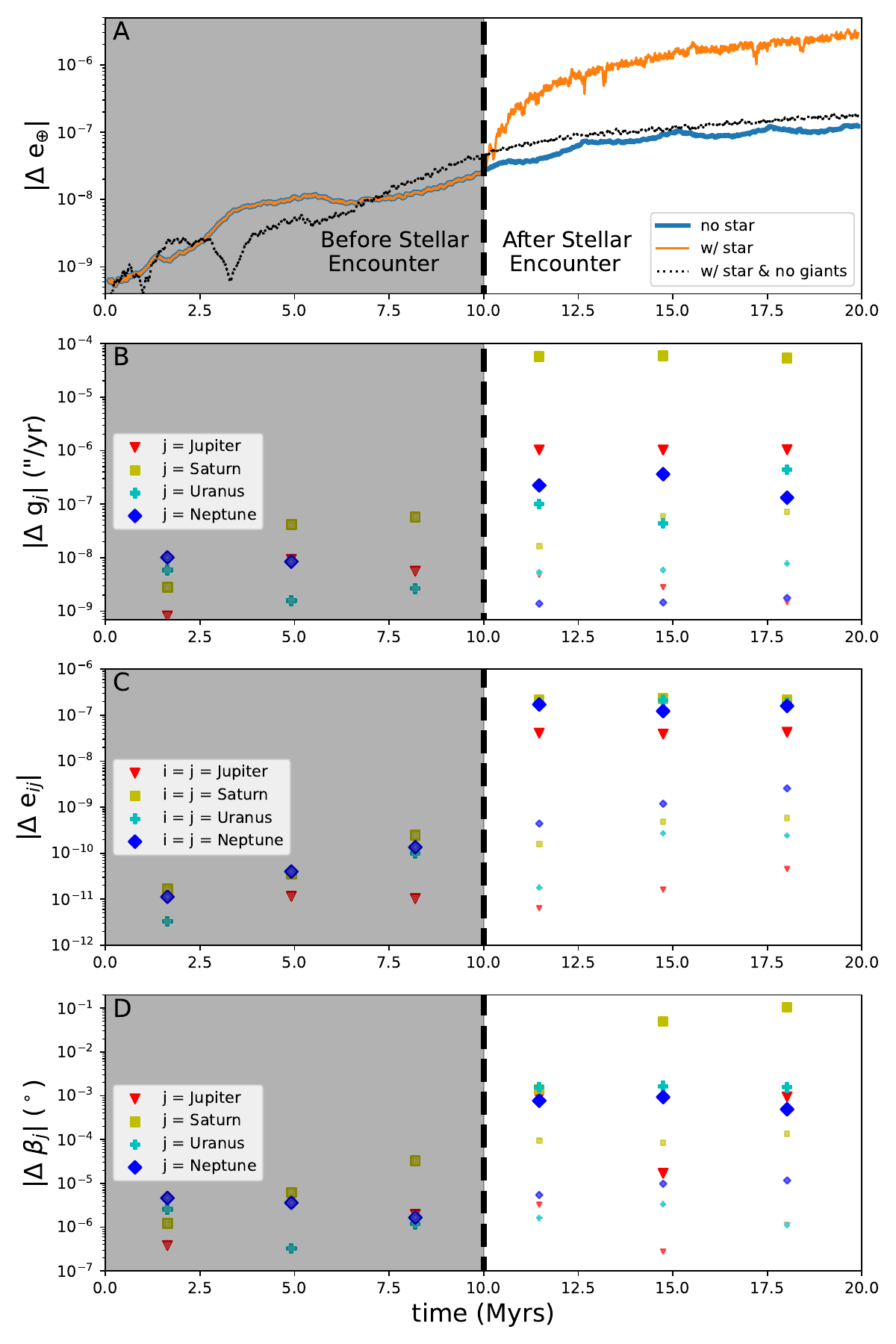}
\caption{{\bf A:} The absolute difference in Earth's orbital eccentricity between pairs of solar system clone integrations is plotted against time. One pair is integrated in complete isolation ({\it blue line}). In the other pairs ({\it orange and black}), one pair member includes the 25 km/s passage of a 1.4 M$_{\odot}$ star within 7200 au of the Sun at $t=10$ Myrs, but the other does not include this passage. One pair ({\it black}) also excludes the giant planets. {\bf B:} For the clone pairs including the giant planets, the absolute pair difference in the proper secular eccentricity precession frequency for each giant planet ($g_j$ in Equations 1 and 2) is shown vs time. Planets from the isolated clone pair ({\it smaller, lighter symbols}) and the stellar-perturbed pair ({\it larger, darker symbols}) are shown. {\bf C:} For the clone pairs including the giant planets, the absolute pair difference in the amplitude of the proper secular eccentricity precession frequency for each giant planet is shown vs time. Planets from the isolated clone pair ({\it smaller, lighter symbols}) and the stellar-perturbed pair ({\it larger, darker symbols}) are shown. {\bf D:} For the clone pairs including the giant planets, the absolute pair difference in the phase of the proper secular eccentricity precession frequency for each giant planet is shown vs time. Planets from the isolated clone pair ({\it smaller, lighter symbols}) and the stellar-perturbed pair ({\it larger, darker symbols}) are shown.}
\label{fig:example}
\end{figure}

The mechanism by which stars alter Earth's orbital evolution is illustrated in Figure \ref{fig:example}. In the top panel, we plot the difference in Earth's orbital eccentricity for pairs of solar system integrations with minutely different initial conditions ($\pm2$ cm). Among a pair that is integrated in isolation, Earth attains an eccentricity difference of $\sim$10$^{-7}$ after 20 Myrs of evolution. However, when one of the pair members is subjected to a 25 km/s passage of a 1.4 M$_{\odot}$ star within 7200 au, the eccentricity difference jumps up by over an order of magnitude to $\sim$$3\times10^{-6}$ after 20 Myrs of integration. The rapid increase in eccentricity divergence after the stellar encounter at 10 Myrs is obvious. In a final pair integration, we still expose one of the members to the stellar encounter, but this time the giant planets are not included in either simulation. In this last case, the effect of the stellar encounter disappears from the plot, and the pair finishes with an eccentricity difference very near that of the original isolated pair. This strongly suggests that the giant planets play an important role in connecting Earth's orbital evolution to stellar passages. 

Even in isolation, the giant planets play a critical role in Earth's orbital evolution (as well as the other inner planets' orbits). Any planet's eccentricity evolution with time can be linearly approximated within Lagrange-Laplace theory as shown in Equations 1 \& 2.  In this expression, $e_i$ and $\varpi_i$ are the eccentricity and longitude of perihelion of the $i$th planet from the Sun, while $g_j$ and $\beta_j$ are the eigenfrequency and phase at which planet $j$'s eccentricity vector precesses, and $M_{ij}$ can be thought of as the magnitude of the eccentricity forcing planet $j$ exerts on planet $i$. (An analogous decomposition can also be performed for a planet's inclination nodal regression frequencies, $s_j$.) In Earth's case ($i=3$), the magnitude of Jupiter's forcing $M_{3,5}$ is $\sim$0.019, or more than half of Earth's mean orbital eccentricity. Thus, the instantaneous value of Earth's orbital eccentricity is highly dependent on Jupiter's orbital evolution (which in turn can be influenced by the other giant planets). \\

\begin{equation}
e_i \cos{\varpi_i} = \sum_{j=1}^8 M_{ij}\cos{(g_jt + \beta_j)}
\end{equation}
\begin{equation}
e_i \sin{\varpi_i} = \sum_{j=1}^8 M_{ij}\sin{(g_jt + \beta_j)}
\end{equation}

Figures 2B, 2C, and 2D illustrate the impact that a stellar passage can have on the eigenfrequencies, phases, and magnitudes of the giant planets' eccentricity forcing. In Figure \ref{fig:example}B we plot the change in $g_j$ over time between panel A's clone pair in which one member is subjected to a stellar passage and the other is not. We also do this for the clone pair integrated in isolation. As can be seen, for the first 10 Myrs, the two pairs of systems have identical drifts in eigenfrequency for all four gas giants. However, the divergence in eigenfrequency becomes 2--3 orders greater in the stellar-perturbed pair after the stellar encounter occurs. 

Figure \ref{fig:example}C shows that the stellar passage has a similar effect on the change in $M_{ii}$. Within the context of Lagrange-Laplace theory, $M_{ii}$ is effectively the eccentricity that the $i$th planet would have in the absence of all other planetary perturbations. Here we see that the stellar encounter drives an $M_{ii}$ divergence among the giant planets that is 2--4 orders of magnitude greater when one system is perturbed by the stellar passage compared to a pair of completely isolated systems. Finally, Figure \ref{fig:example}D shows that the divergence in the phases of the giant planets' eccentricity forcing is 1--3 orders of magnitudes greater when a stellar encounter perturbs a pair member than when both are isolated. Thus, the frequencies, magnitudes, and phases at which the giant planets perturb the inner planets change much more quickly in the presence of stellar passages than in isolation. In this way, the giant planets serve as a dynamical link that ultimately allows the Milky Way's stars to influence the long-term evolution of Earth's orbit.

\section{Parameterizing Stellar Passage Importance} \label{sec:floats}

The degree to which the giant planets' orbits (and by extension the Earth's orbit) are perturbed of course depends on the strength of the stellar perturbation. One way to measure stellar passage strength is by the velocity impulse it imparts upon the Sun, or $\frac{2GM_{*}}{v_{*}b_{\odot}}$, where $b_{\odot}$ is the stellar impact parameter with respect to the Sun, $v_{*}$ is the passage velocity, and $M_{*}$ is the stellar mass \citep{rick76}. If a planet's dynamical timescale is much longer than the passage timescale (known as the impulsive regime), the planet's impulse relative to the Sun's is then $\frac{2GM_{*}}{v_{*}b_{\odot}^2}\Delta x$, where $\Delta x$ is the relative separation of the Sun and planet projected along the perpendicular to the star's path at its nearest approach to the Sun. 

\begin{figure}
\centering
\includegraphics[width=.8\linewidth]{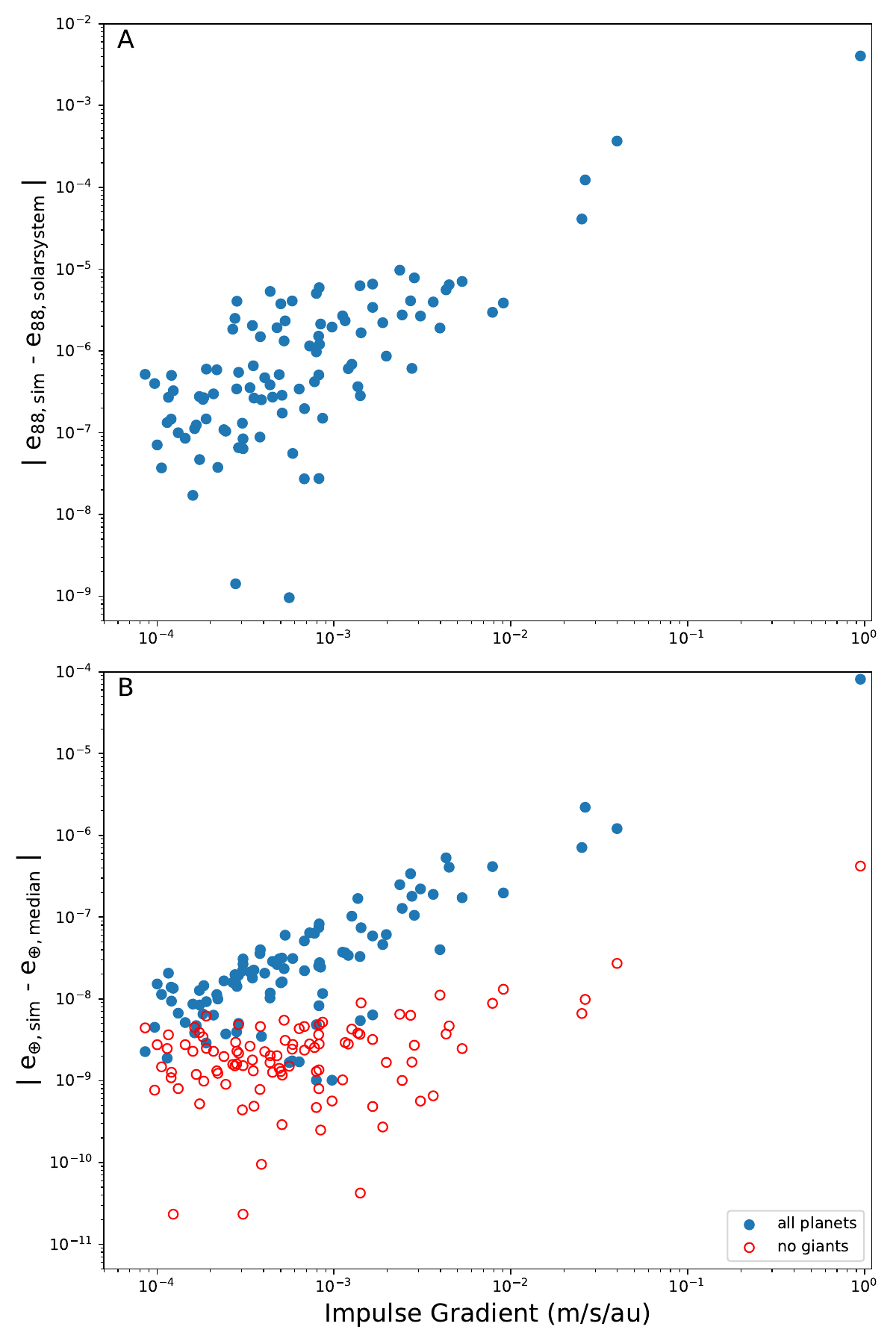}
\caption{{\bf A:} The absolute difference between the observed amplitude of Neptune's proper secular eccentricity precession frequency and that measured in 100 integrations of solar system clones is plotted against the impulse gradient of the stellar encounter to which each clone is exposed.  {\bf B:} The absolute difference between the Earth's eccentricity in 100 solar system clone integrations and the median eccentricity is plotted against the impulse gradient of the stellar encounter to which each clone is exposed. Filled blue data points are for clones that contain all 8 planets, while open red data points do not contain the giant planets. Earth eccentricities are measured $\sim$1 Myr after the stellar encounter in each integration.}
\label{fig:impgrad}
\end{figure}

In fact, most of our stellar passage timescales are far from the impulsive regime. The opposite limit, in which the encounter timescale is much longer than the planetary dynamical timescale, is known as the adiabatic regime.  Typically, the planetary dynamical timescale is taken to be the orbital period. However, since we are primarily studying passing stars' perturbations on the planets' secular architecture (see Figure \ref{fig:example}), the most relevant planetary timescale is likely the planetary precession periods, which are $\sim$10$^{5-6}$ years. Thus, with respect to perturbations on the solar system's secular architecture, many stellar passages will be non-adiabatic. 

Although most of our stellar passages are far from the formal impulsive regime, we nonetheless find that the perturbation delivered to the solar system scales with $\frac{2GM_{*}}{v_{*}b_{\odot}^2}$, which is effectively the impulse gradient along the perpendicular to the passage path at nearest approach \citep{ray23}. This is illustrated in Figure \ref{fig:impgrad}A. Here we have generated 100 different 50-Myr sequences of stellar encounters. In each sequence, we select only the encounter with the largest impulse gradient and then subject a solar system clone to this encounter. The perturbation delivered to $M_{8,8}$ (approximately Neptune's eccentricity in the absence of other planetary forcing) is then plotted against the impulse gradient of these 100 different encounters. The two quantities are well-correlated across 4--5 orders of magnitude of perturbation. Perturbations to other giant planet orbital parameters display a level of correlation with stellar impulse gradient that is similar to $M_{8,8}$. 

In Figure \ref{fig:impgrad}B, we see that these giant planet perturbations ultimately lead to perturbations on Earth's orbit. Measuring Earth's eccentricity 1 Myr after the stellar passages, we find that the deviation of Earth's eccentricity from the ensemble's median eccentricity scales with the stellar impulse gradient in a manner similar to the giant planets. To once again highlight the importance of the giant planets' role, we repeat these stellar passage experiments without the giant planets. In these repeat runs, we see that the perturbation to Earth's orbit falls by $\sim$2 orders of magnitude. This again confirms that Earth's sensitivity to stellar passages primarily relies on the passages' influence over the more distant giant planets. In the event that the solar system were to have even more distant planets, the dynamical impact of stellar passages could be still stronger than what is described here \citep{trushep14, batbrown16}.

\section{HD 7977 and the Paleoclimate} \label{subsec:tables}

It is clear that the stellar passages expected within the solar neighborhood significantly influence the orbital evolution of the Sun's planets, and we now assess the effects of a specific encounter known to have occurred. Amongst past stellar encounters inferred from Gaia Data Release 3, HD 7977 stands out as potentially the closest recent known encounter. This 1.1 M$_{\odot}$ star passed near the solar system $\sim$2.8 Myrs ago at $\sim$27 km/s \citep{bail22, gaia23}. Although this encounter's median inferred impact parameter is $\sim$13200 au, there is a large amount of uncertainty, with a 5\% probability of passage within $\sim$3900 au. This range of impact parameter corresponds to over an order of magnitude variation in impulse gradient (which governs the level of planetary perturbation). 

\begin{figure}
\centering
\includegraphics[width=.8\linewidth]{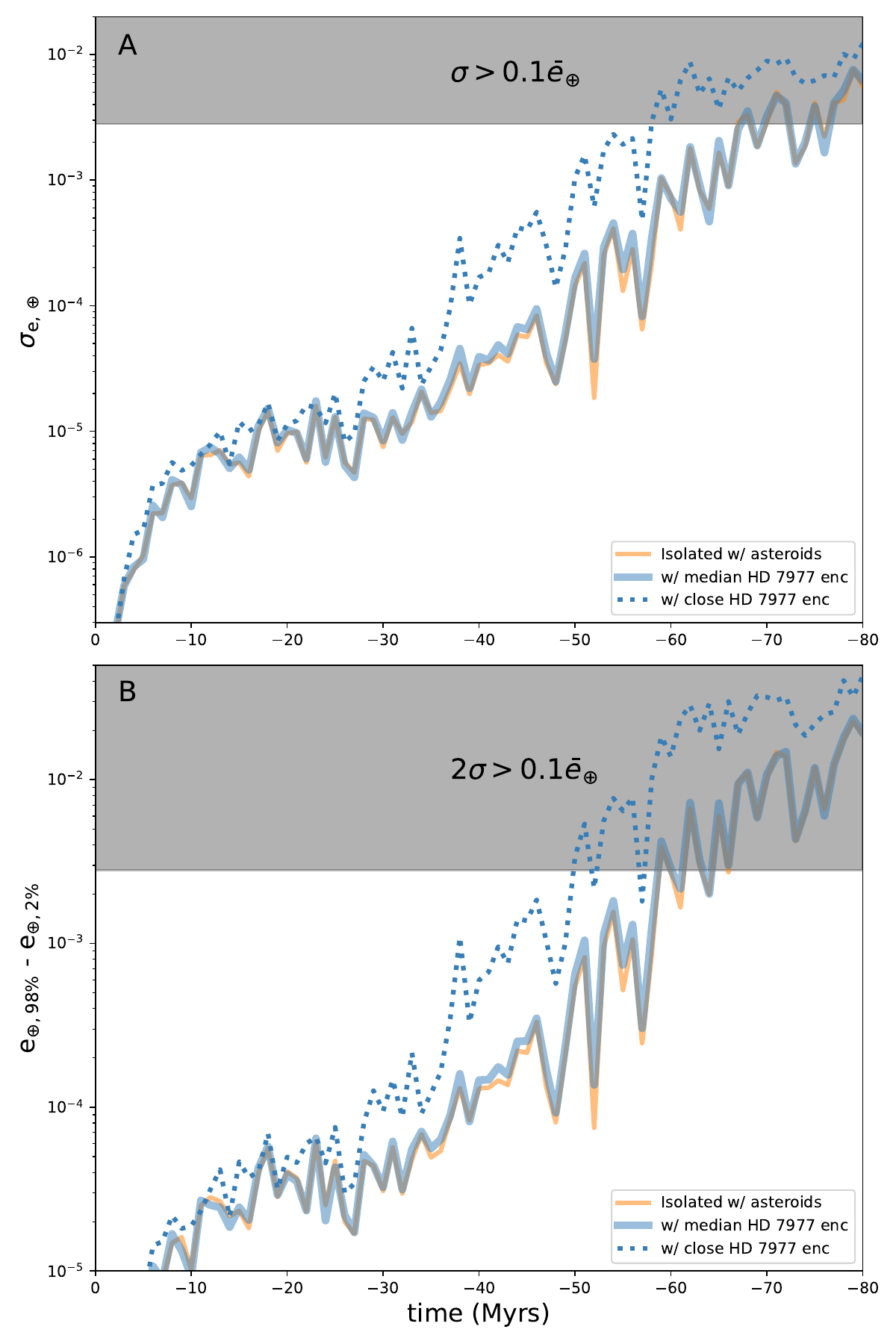}
\caption{{\bf A:} The standard deviation of Earth's orbital eccentricity is plotted against time for 100 back-integrations of solar system clones. Three batches of solar system simulations include the Sun, its eight planets, Pluto, Ceres, Vesta, Pallas, Iris, and Bamberga. One batch includes no stellar perturbations ({\it orange}). Another also includes a passage of HD 7977 at $t=-2.8$ Myrs at its median impact parameter of 13200 au, as indicated by Gaia uncertainties ({\it solid blue}) \citep{bail22}. The final batch includes a passage of HD 7977 at $t=-2.8$ Myrs at 3900 au, the 5th percentile of possible impact parameters according to Gaia uncertainties ({\it dashed blue}) \citep{bail22}. {\bf B:} The difference between the 98th and 2nd percentile of Earth's orbital eccentricity vs time among 100 back-integrations of solar system clones. The three batches of clones are the same as in the top panel.}
\label{fig:HDeccsigma}
\end{figure}

In Figure \ref{fig:HDeccsigma}A, we back-integrate 100 solar system realizations analogous to the isolated ensemble including asteroids from Figure \ref{fig:eccsigma}. However, this time we also subject the ensemble to a HD 7977-like passage at $t=-2.8$ Myrs. This is done twice. In the first instance, we hold the impact parameter at the median value of 13200 au, while our individual clones randomly vary the encounter velocity ($\pm0.4$ km/s) and time of passage ($\pm0.004$ Myrs) within the observational uncertainty range \citep{bail22}. The second back-integration ensemble employs the same variations in stellar velocity and passage time, but the impact parameter is decreased to 3900 au. As seen in Figure \ref{fig:HDeccsigma}A, the median impact parameter for HD 7977 has no discernible effect on the divergence of Earth's eccentricity during back-integrations. Given that the real encounter's impact parameter was probably near this value or larger, HD 7977 has most likely not had a significant effect on the solar system's long-term chaotic evolution. However, this is not for certain. We find that the stellar passage's effect becomes very significant if we assume the 3900 au impact parameter, which has a 5\% likelihood. In this case, the standard deviation of Earth's eccentricity exceeds 0.0028 $\sim$10 Myrs faster ($t=-58$ Myrs) than our models that just consider planets and asteroids.  

In this last back-integration ensemble that considers the smaller impact parameter with 5\% likelihood, even faster divergence is also possible. If we instead plot the difference between the 98th and 2nd percentile of Earth's eccentricity (roughly the 2$\sigma$ confidence bounds), we see that the difference exceeds 0.0028 in just 50 Myrs, nearly 20 Myrs faster than the nominal chaos time horizon set by asteroid interactions in our models. We should note that employing a 3900 au impact parameter for HD 7977 would likely make it one of the ten most powerful encounters experienced over the solar system's history as measured by impulse gradient (owing to the star's above average mass and below average encounter velocity). Nevertheless, such an impact parameter is within the bounds of observational uncertainty, and better constraining this encounter is essential to understanding the recent dynamical history of Earth and the rest of the Sun's planets.

Incredibly precise dynamical models of Earth's past orbital evolution have been created over the last 3 decades to assist interpreting the past 50--70 Myrs of Earth's sedimentation record and paleoclimate \citep{lask04, lask11b, zeeb19}. However, beyond 50 Myrs, these model outcomes are known to be sensitive to slight changes in solar system properties such as the Sun's oblateness and the number of asteroids considered \citep{zeeb17}, and models that maximize agreement between Earth's eccentricity evolution and sedimentation patterns 50--60 Myrs ago must invoke modern solar system parameters that are somewhat in tension with observational constraints \citep{zeeb19}. These maximal agreement models also contain a specific resonance transition 50 Myrs ago \citep{lask04, ma17, zeeb19}. Currently, the eigenfrequency difference between Earth's and Mars' nodal regression ($s_4 - s_3$) is twice that of their perihelion precession difference ($g_4 - g_3$), and entry into this resonance occurred when the perihelion precession converged to a ratio of 2 from some other value at some point in the past. 

\begin{figure}
\centering
\includegraphics[width=.8\linewidth]{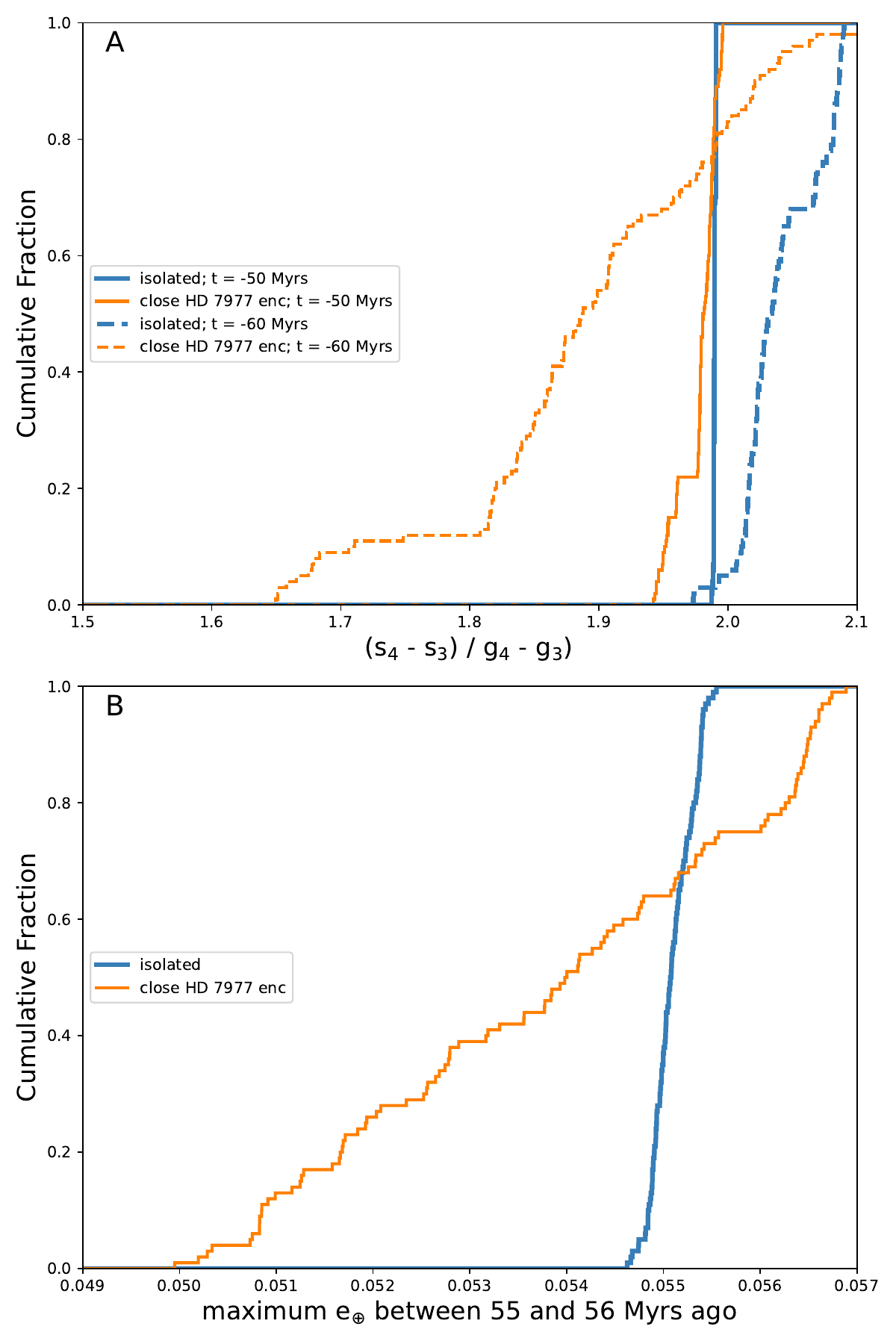}
\caption{{\bf A:} The distribution of $(s_4-s_3)/(g_4-g_3)$ is calculated for our isolated back-integrated systems ({\it blue}) and our back-integrated systems perturbed by a 3900 au flyby of HD 7977 at $t=-2.8$ Myrs ({\it orange}). These ratios are calculated at $t=-50$ Myrs ({\it solid}) and $t=-60$ Myrs ({\it dashed}). {\bf B:} The distribution of maximum eccentricity values Earth attains between $-55$ Myrs and $-56$ Myrs is plotted for our isolated solar system realizations ({\it blue}) and our systems perturbed by a 3900 au flyby of HD 7977 at $t=-2.8$ Myrs ({\it orange}).}
\label{fig:restrans}
\end{figure}

In Figure \ref{fig:restrans}A, we calculate the ratio ($s_4 - s_3$) to ($g_4 - g_3$) among our solar system realizations from Figure \ref{fig:HDeccsigma} at $t=-50$ Myrs. We see that none of our isolated systems are far from the 2:1 resonance at this past time. Meanwhile, the systems subjected to HD 7977's 3900 au flyby are also concentrated near 2:1 but display larger variance. When we re-examine these ratios at the earlier epoch of $t=-60$ Myrs, we find that our isolated systems are systematically nearer to 2:1 than the HD 7977-perturbed systems, many of which are clearly well outside the 2:1 resonance and have yet to enter it. Thus, the close encounter with HD 7977 at $t=-2.8$ Myrs can unlock a much broader possible spectrum of orbital behavior 50--60 Myrs ago with respect to entry into this secular resonance. The consequences of this can be seen in Figure \ref{fig:restrans}B, where we examine the maximum eccentricity Earth attains 55--56 Myrs ago. This highlighted period encompasses the Paleocene--Eocene Thermal Maximum, which has been suggested to coincide with a period of heightened orbital eccentricity for Earth \citep{zeeb19}. Among our 100 isolated solar system realizations, Earth's maximum eccentricity during this time varies between 0.0546 and 0.0555 ($\pm 0.00045$). Among our 100 HD 7977-perturbed systems, the range in maximum eccentricity is nearly an order of magnitude larger at 0.0500--0.0569 ($\pm0.0035$). Thus, our interpretation of Earth's orbit near the Paleocene--Eocene Thermal Maximum is dependent on the exact parameters of stellar encounters that have occurred in the past several Myrs. Due to the broader range of Earth's possible orbital behavior, orbital solutions may exist that provide good matches to the geological record before 50 Myrs ago but also do not necessarily have a period of notably heightened orbital eccentricity during the Paleocene--Eocene Thermal Maximum. Conversely, the divergent behaviors illustrated in Figure \ref{fig:restrans}A raise the possibility that geological constraints on the timing of this secular resonance entry could imply additional constraints on the magnitude of recent stellar perturbations \citep{west17, ma17}. 

\section{Summary \& Conclusions}

Much more sophisticated simulations of Earth's past orbital evolution exist compared to those presented here \citep{lask04, lask11b, zeeb19}, and our present work is not meant to make the most accurate predictions of Earth's exact orbital history. Instead, it compares the importance of stellar encounters relative to the much more intensely studied internal dynamics of the solar system. We show that stellar encounters play an important role in our solar system's long-term dynamical evolution, and our results highlight a number of notable conclusions. First, stellar encounters significantly accelerate the chaotic diffusion of Earth's orbit, and the time back to which numerical simulations can confidently predict Earth's orbital evolution is $\sim$10\% shorter than previously thought. Second, this chaotic divergence that stellar passages impart on Earth's orbit results from their perturbations to the giant planets' orbits, and these perturbations roughly scale with the velocity impulse gradients of stellar encounters. Third, the known encounter with HD 7977 2.8 Myrs ago has the potential to unlock new sequences of Earth's past orbital evolution beyond 50 Myrs ago that have not been considered or generated in previous modeling efforts. Although it takes tens of Myrs for the effects of stellar passages to significantly manifest themselves, the long-term orbital evolution of the Earth and the rest of the planets is linked to these stars.

\section{Acknowledgements}

This work was performed with support from NSF CAREER Award 2405121. SNR acknowledges support from the French Programme Nationale de Planetologie (PNP) and in the framework of the Investments for the Future programme IdEx, Universit\'e de Bordeaux/RRI ORIGINS. Computing for this project was performed at the OU Supercomputing Center for Education \& Research (OSCER) at the University of Oklahoma (OU). 

\bibliography{scibib}{}

\newcommand{\noop}[1]{}
\begin{thebibliography}{}
\expandafter\ifx\csname natexlab\endcsname\relax\def\natexlab#1{#1}\fi
\providecommand{\url}[1]{\href{#1}{#1}}
\providecommand{\dodoi}[1]{doi:~\href{http://doi.org/#1}{\nolinkurl{#1}}}
\providecommand{\doeprint}[1]{\href{http://ascl.net/#1}{\nolinkurl{http://ascl.net/#1}}}
\providecommand{\doarXiv}[1]{\href{https://arxiv.org/abs/#1}{\nolinkurl{https://arxiv.org/abs/#1}}}

\bibitem[{{Bailer-Jones}(2018)}]{bail18}
{Bailer-Jones}, C.~A.~L. 2018, \it Astronomy \& Astrophysics, 609, A8,
  \dodoi{10.1051/0004-6361/201731453}

\bibitem[{{Bailer-Jones}(2022)}]{bail22}
---. 2022, The Astrophysical Journal Letters, 935, L9,
  \dodoi{10.3847/2041-8213/ac816a}

\bibitem[{{Batygin} \& {Brown}(2016)}]{batbrown16}
{Batygin}, K., \& {Brown}, M.~E. 2016, \aj, 151, 22,
  \dodoi{10.3847/0004-6256/151/2/22}

\bibitem[{{Brown} \& {Rein}(2022)}]{brownrein22}
{Brown}, G., \& {Rein}, H. 2022, \it Monthly Notices of the Royal Astronomical
  Society, 515, 5942, \dodoi{10.1093/mnras/stac1763}

\bibitem[{{Brunini}(1993)}]{brun93}
{Brunini}, A. 1993, Astronomy \& Astrophysics, 276, 261

\bibitem[{{Chambers}(1999)}]{cham99}
{Chambers}, J.~E. 1999, Monthly Notices of the Royal Astronomical Society, 304,
  793, \dodoi{10.1046/j.1365-8711.1999.02379.x}

\bibitem[{{Gaia Collaboration} {et~al.}(2023){Gaia Collaboration}, {Vallenari},
  {Brown}, {Prusti}, {de Bruijne}, {Arenou}, {Babusiaux}, {Biermann},
  {Creevey}, {Ducourant}, {Evans}, {Eyer}, {Guerra}, {Hutton}, {Jordi},
  {Klioner}, {Lammers}, {Lindegren}, {Luri}, {Mignard}, {Panem}, {Pourbaix},
  {Randich}, {Sartoretti}, {Soubiran}, {Tanga}, {Walton}, {Bailer-Jones},
  {Bastian}, {Drimmel}, {Jansen}, {Katz}, {Lattanzi}, {van Leeuwen}, {Bakker},
  {Cacciari}, {Casta{\~n}eda}, {De Angeli}, {Fabricius}, {Fouesneau},
  {Fr{\'e}mat}, {Galluccio}, {Guerrier}, {Heiter}, {Masana}, {Messineo},
  {Mowlavi}, {Nicolas}, {Nienartowicz}, {Pailler}, {Panuzzo}, {Riclet}, {Roux},
  {Seabroke}, {Sordo}, {Th{\'e}venin}, {Gracia-Abril}, {Portell}, {Teyssier},
  {Altmann}, {Andrae}, {Audard}, {Bellas-Velidis}, {Benson}, {Berthier},
  {Blomme}, {Burgess}, {Busonero}, {Busso}, {C{\'a}novas}, {Carry}, {Cellino},
  {Cheek}, {Clementini}, {Damerdji}, {Davidson}, {de Teodoro}, {Nu{\~n}ez
  Campos}, {Delchambre}, {Dell'Oro}, {Esquej}, {Fern{\'a}ndez-Hern{\'a}ndez},
  {Fraile}, {Garabato}, {Garc{\'\i}a-Lario}, {Gosset}, {Haigron}, {Halbwachs},
  {Hambly}, {Harrison}, {Hern{\'a}ndez}, {Hestroffer}, {Hodgkin}, {Holl},
  {Jan{\ss}en}, {Jevardat de Fombelle}, {Jordan}, {Krone-Martins}, {Lanzafame},
  {L{\"o}ffler}, {Marchal}, {Marrese}, {Moitinho}, {Muinonen}, {Osborne},
  {Pancino}, {Pauwels}, {Recio-Blanco}, {Reyl{\'e}}, {Riello}, {Rimoldini},
  {Roegiers}, {Rybizki}, {Sarro}, {Siopis}, {Smith}, {Sozzetti}, {Utrilla},
  {van Leeuwen}, {Abbas}, {{\'A}brah{\'a}m}, {Abreu Aramburu}, {Aerts},
  {Aguado}, {Ajaj}, {Aldea-Montero}, {Altavilla}, {{\'A}lvarez}, {Alves},
  {Anders}, {Anderson}, {Anglada Varela}, {Antoja}, {Baines}, {Baker},
  {Balaguer-N{\'u}{\~n}ez}, {Balbinot}, {Balog}, {Barache}, {Barbato},
  {Barros}, {Barstow}, {Bartolom{\'e}}, {Bassilana}, {Bauchet}, {Becciani},
  {Bellazzini}, {Berihuete}, {Bernet}, {Bertone}, {Bianchi}, {Binnenfeld},
  {Blanco-Cuaresma}, {Blazere}, {Boch}, {Bombrun}, {Bossini}, {Bouquillon},
  {Bragaglia}, {Bramante}, {Breedt}, {Bressan}, {Brouillet}, {Brugaletta},
  {Bucciarelli}, {Burlacu}, {Butkevich}, {Buzzi}, {Caffau}, {Cancelliere},
  {Cantat-Gaudin}, {Carballo}, {Carlucci}, {Carnerero}, {Carrasco},
  {Casamiquela}, {Castellani}, {Castro-Ginard}, {Chaoul}, {Charlot}, {Chemin},
  {Chiaramida}, {Chiavassa}, {Chornay}, {Comoretto}, {Contursi}, {Cooper},
  {Cornez}, {Cowell}, {Crifo}, {Cropper}, {Crosta}, {Crowley}, {Dafonte},
  {Dapergolas}, {David}, {David}, {de Laverny}, {De Luise}, {De March}, {De
  Ridder}, {de Souza}, {de Torres}, {del Peloso}, {del Pozo}, {Delbo},
  {Delgado}, {Delisle}, {Demouchy}, {Dharmawardena}, {Di Matteo}, {Diakite},
  {Diener}, {Distefano}, {Dolding}, {Edvardsson}, {Enke}, {Fabre}, {Fabrizio},
  {Faigler}, {Fedorets}, {Fernique}, {Fienga}, {Figueras}, {Fournier},
  {Fouron}, {Fragkoudi}, {Gai}, {Garcia-Gutierrez}, {Garcia-Reinaldos},
  {Garc{\'\i}a-Torres}, {Garofalo}, {Gavel}, {Gavras}, {Gerlach}, {Geyer},
  {Giacobbe}, {Gilmore}, {Girona}, {Giuffrida}, {Gomel}, {Gomez},
  {Gonz{\'a}lez-N{\'u}{\~n}ez}, {Gonz{\'a}lez-Santamar{\'\i}a},
  {Gonz{\'a}lez-Vidal}, {Granvik}, {Guillout}, {Guiraud},
  {Guti{\'e}rrez-S{\'a}nchez}, {Guy}, {Hatzidimitriou}, {Hauser}, {Haywood},
  {Helmer}, {Helmi}, {Sarmiento}, {Hidalgo}, {Hilger}, {H{\l}adczuk}, {Hobbs},
  {Holland}, {Huckle}, {Jardine}, {Jasniewicz}, {Jean-Antoine Piccolo},
  {Jim{\'e}nez-Arranz}, {Jorissen}, {Juaristi Campillo}, {Julbe}, {Karbevska},
  {Kervella}, {Khanna}, {Kontizas}, {Kordopatis}, {Korn}, {K{\'o}sp{\'a}l},
  {Kostrzewa-Rutkowska}, {Kruszy{\'n}ska}, {Kun}, {Laizeau}, {Lambert},
  {Lanza}, {Lasne}, {Le Campion}, {Lebreton}, {Lebzelter}, {Leccia}, {Leclerc},
  {Lecoeur-Taibi}, {Liao}, {Licata}, {Lindstr{\o}m}, {Lister}, {Livanou},
  {Lobel}, {Lorca}, {Loup}, {Madrero Pardo}, {Magdaleno Romeo}, {Managau},
  {Mann}, {Manteiga}, {Marchant}, {Marconi}, {Marcos}, {Marcos Santos},
  {Mar{\'\i}n Pina}, {Marinoni}, {Marocco}, {Marshall}, {Martin Polo},
  {Mart{\'\i}n-Fleitas}, {Marton}, {Mary}, {Masip}, {Massari},
  {Mastrobuono-Battisti}, {Mazeh}, {McMillan}, {Messina}, {Michalik}, {Millar},
  {Mints}, {Molina}, {Molinaro}, {Moln{\'a}r}, {Monari}, {Mongui{\'o}},
  {Montegriffo}, {Montero}, {Mor}, {Mora}, {Morbidelli}, {Morel}, {Morris},
  {Muraveva}, {Murphy}, {Musella}, {Nagy}, {Noval}, {Oca{\~n}a}, {Ogden},
  {Ordenovic}, {Osinde}, {Pagani}, {Pagano}, {Palaversa}, {Palicio},
  {Pallas-Quintela}, {Panahi}, {Payne-Wardenaar}, {Pe{\~n}alosa Esteller},
  {Penttil{\"a}}, {Pichon}, {Piersimoni}, {Pineau}, {Plachy}, {Plum}, {Poggio},
  {Pr{\v{s}}a}, {Pulone}, {Racero}, {Ragaini}, {Rainer}, {Raiteri}, {Rambaux},
  {Ramos}, {Ramos-Lerate}, {Re Fiorentin}, {Regibo}, {Richards}, {Rios Diaz},
  {Ripepi}, {Riva}, {Rix}, {Rixon}, {Robichon}, {Robin}, {Robin}, {Roelens},
  {Rogues}, {Rohrbasser}, {Romero-G{\'o}mez}, {Rowell}, {Royer}, {Ruz Mieres},
  {Rybicki}, {Sadowski}, {S{\'a}ez N{\'u}{\~n}ez}, {Sagrist{\`a} Sell{\'e}s},
  {Sahlmann}, {Salguero}, {Samaras}, {Sanchez Gimenez}, {Sanna},
  {Santove{\~n}a}, {Sarasso}, {Schultheis}, {Sciacca}, {Segol}, {Segovia},
  {S{\'e}gransan}, {Semeux}, {Shahaf}, {Siddiqui}, {Siebert}, {Siltala},
  {Silvelo}, {Slezak}, {Slezak}, {Smart}, {Snaith}, {Solano}, {Solitro},
  {Souami}, {Souchay}, {Spagna}, {Spina}, {Spoto}, {Steele},
  {Steidelm{\"u}ller}, {Stephenson}, {S{\"u}veges}, {Surdej}, {Szabados},
  {Szegedi-Elek}, {Taris}, {Taylor}, {Teixeira}, {Tolomei}, {Tonello}, {Torra},
  {Torra}, {Torralba Elipe}, {Trabucchi}, {Tsounis}, {Turon}, {Ulla}, {Unger},
  {Vaillant}, {van Dillen}, {van Reeven}, {Vanel}, {Vecchiato}, {Viala},
  {Vicente}, {Voutsinas}, {Weiler}, {Wevers}, {Wyrzykowski}, {Yoldas}, {Yvard},
  {Zhao}, {Zorec}, {Zucker}, \& {Zwitter}}]{gaia23}
{Gaia Collaboration}, {Vallenari}, A., {Brown}, A.~G.~A., {et~al.} 2023, \aap,
  674, A1, \dodoi{10.1051/0004-6361/202243940}

\bibitem[{{Garc{\'\i}a-S{\'a}nchez}
  {et~al.}(2001{\natexlab{a}}){Garc{\'\i}a-S{\'a}nchez}, {Weissman}, {Preston},
  {Jones}, {Lestrade}, {Latham}, {Stefanik}, \& {Paredes}}]{gar01}
{Garc{\'\i}a-S{\'a}nchez}, J., {Weissman}, P.~R., {Preston}, R.~A., {et~al.}
  2001{\natexlab{a}}, Astronomy \& Astrophysics, 379, 634,
  \dodoi{10.1051/0004-6361:20011330}

\bibitem[{{Garc{\'\i}a-S{\'a}nchez}
  {et~al.}(2001{\natexlab{b}}){Garc{\'\i}a-S{\'a}nchez}, {Weissman}, {Preston},
  {Jones}, {Lestrade}, {Latham}, {Stefanik}, \& {Paredes}}]{garc01}
---. 2001{\natexlab{b}}, Astronomy \& Astrophysics, 379, 634,
  \dodoi{10.1051/0004-6361:20011330}

\bibitem[{{Hays} {et~al.}(1976){Hays}, {Imbrie}, \& {Shackleton}}]{hays76}
{Hays}, J.~D., {Imbrie}, J., \& {Shackleton}, N.~J. 1976, \it Science, 194,
  1121, \dodoi{10.1126/science.194.4270.1121}

\bibitem[{{Henon}(1972)}]{hen72}
{Henon}, M. 1972, Astronomy \& Astrophysics, 19, 488

\bibitem[{{Kaib} {et~al.}(2018){Kaib}, {White}, \& {Izidoro}}]{kaib18}
{Kaib}, N.~A., {White}, E.~B., \& {Izidoro}, A. 2018, Monthly Notices of the
  Royal Astronomical Society, 473, 470, \dodoi{10.1093/mnras/stx2456}

\bibitem[{{Laskar}(1988)}]{lask88}
{Laskar}, J. 1988, \it Astronomy \& Astrophysics, 198, 341

\bibitem[{{Laskar}(1989)}]{lask89}
---. 1989, \it Nature, 338, 237, \dodoi{10.1038/338237a0}

\bibitem[{{Laskar}(1990)}]{lask90}
---. 1990, \it Icarus, 88, 266, \dodoi{10.1016/0019-1035(90)90084-M}

\bibitem[{{Laskar}(1994)}]{lask94}
---. 1994, \it Astronomy \& Astrophysics, 287, L9

\bibitem[{{Laskar}(1999)}]{lask99}
---. 1999, Philosophical Transactions of the Royal Society of London Series A,
  357, 1735, \dodoi{10.1098/rsta.1999.0399}

\bibitem[{{Laskar}(2008)}]{lask08}
---. 2008, \it Icarus, 196, 1, \dodoi{10.1016/j.icarus.2008.02.017}

\bibitem[{{Laskar} {et~al.}(2011{\natexlab{a}}){Laskar}, {Fienga}, {Gastineau},
  \& {Manche}}]{lask11b}
{Laskar}, J., {Fienga}, A., {Gastineau}, M., \& {Manche}, H.
  2011{\natexlab{a}}, Astronomy \& Astrophysics, 532, A89,
  \dodoi{10.1051/0004-6361/201116836}

\bibitem[{{Laskar} {et~al.}(2011{\natexlab{b}}){Laskar}, {Gastineau},
  {Delisle}, {Farr{\'e}s}, \& {Fienga}}]{lask11}
{Laskar}, J., {Gastineau}, M., {Delisle}, J.~B., {Farr{\'e}s}, A., \& {Fienga},
  A. 2011{\natexlab{b}}, \it Astrononmy \& Astrophysics, 532, L4,
  \dodoi{10.1051/0004-6361/201117504}

\bibitem[{{Laskar} {et~al.}(2004){Laskar}, {Robutel}, {Joutel}, {Gastineau},
  {Correia}, \& {Levrard}}]{lask04}
{Laskar}, J., {Robutel}, P., {Joutel}, F., {et~al.} 2004, \it Astronomy \&
  Astrophysics, 428, 261, \dodoi{10.1051/0004-6361:20041335}

\bibitem[{M.(1941)}]{mil41}
M., M.~M. 1941, Koniglich Serbische Akademice Beograd Special Publication, 132.
\newblock \url{https://cir.nii.ac.jp/crid/1571698599073715328}

\bibitem[{{Ma} {et~al.}(2017){Ma}, {Meyers}, \& {Sageman}}]{ma17}
{Ma}, C., {Meyers}, S.~R., \& {Sageman}, B.~B. 2017, Nature, 542, 468,
  \dodoi{10.1038/nature21402}

\bibitem[{{Oort}(1950)}]{oort50}
{Oort}, J.~H. 1950, Bulletin of the Astronomical Institutes of the Netherlands,
  11, 91

\bibitem[{{{\"O}pik}(1932)}]{opik32}
{{\"O}pik}, E. 1932, Proceedings of the American Academy of Arts and Sciences,
  67, 169, \dodoi{10.2307/20022899}

\bibitem[{{Quinn} {et~al.}(1991){Quinn}, {Tremaine}, \& {Duncan}}]{quinn91}
{Quinn}, T.~R., {Tremaine}, S., \& {Duncan}, M. 1991, \it Astronomical Journal,
  101, 2287, \dodoi{10.1086/115850}

\bibitem[{{Raymond} {et~al.}(2024){Raymond}, {Kaib}, {Selsis}, \&
  {Bouy}}]{ray23}
{Raymond}, S.~N., {Kaib}, N.~A., {Selsis}, F., \& {Bouy}, H. 2024, \mnras, 527,
  6126, \dodoi{10.1093/mnras/stad3604}

\bibitem[{{Reid} {et~al.}(2002){Reid}, {Gizis}, \& {Hawley}}]{reid02}
{Reid}, I.~N., {Gizis}, J.~E., \& {Hawley}, S.~L. 2002, Astronomical Journal,
  124, 2721, \dodoi{10.1086/343777}

\bibitem[{{Rickman}(1976)}]{rick76}
{Rickman}, H. 1976, Bulletin of the Astronomical Institutes of Czechoslovakia,
  27, 92

\bibitem[{{Rickman} {et~al.}(2008){Rickman}, {Fouchard}, {Froeschl{\'e}}, \&
  {Valsecchi}}]{rick08}
{Rickman}, H., {Fouchard}, M., {Froeschl{\'e}}, C., \& {Valsecchi}, G.~B. 2008,
  Celestial Mechanics and Dynamical Astronomy, 102, 111,
  \dodoi{10.1007/s10569-008-9140-y}

\bibitem[{{Saha} \& {Tremaine}(1992)}]{sahatre92}
{Saha}, P., \& {Tremaine}, S. 1992, Astronomical Journal, 104, 1633,
  \dodoi{10.1086/116347}

\bibitem[{{Sussman} \& {Wisdom}(1988)}]{susswis88}
{Sussman}, G.~J., \& {Wisdom}, J. 1988, \it Science, 241, 433,
  \dodoi{10.1126/science.241.4864.433}

\bibitem[{{Trujillo} \& {Sheppard}(2014)}]{trushep14}
{Trujillo}, C.~A., \& {Sheppard}, S.~S. 2014, \nat, 507, 471,
  \dodoi{10.1038/nature13156}

\bibitem[{{{\v{S}}idlichovsk{\'y}} \& {Nesvorn{\'y}}(1996)}]{sidnes96}
{{\v{S}}idlichovsk{\'y}}, M., \& {Nesvorn{\'y}}, D. 1996, Celestial Mechanics
  and Dynamical Astronomy, 65, 137, \dodoi{10.1007/BF00048443}

\bibitem[{Westerhold {et~al.}(2017)Westerhold, R{\"o}hl, Frederichs, Agnini,
  Raffi, Zachos, \& Wilkens}]{west17}
Westerhold, T., R{\"o}hl, U., Frederichs, T., {et~al.} 2017, Climate of the
  Past, 13, 1129

\bibitem[{{Wisdom} \& {Holman}(1991)}]{wishol91}
{Wisdom}, J., \& {Holman}, M. 1991, \it Astronomical Journal, 102, 1528,
  \dodoi{10.1086/115978}

\bibitem[{{Zeebe}(2015)}]{zeeb15}
{Zeebe}, R.~E. 2015, Astrophysical Journal, 798, 8,
  \dodoi{10.1088/0004-637X/798/1/8}

\bibitem[{{Zeebe}(2017)}]{zeeb17}
---. 2017, Astronomical Journal, 154, 193, \dodoi{10.3847/1538-3881/aa8cce}

\bibitem[{{Zeebe} \& {Lourens}(2019)}]{zeeb19}
{Zeebe}, R.~E., \& {Lourens}, L.~J. 2019, Science, 365, 926,
  \dodoi{10.1126/science.aax0612}

\bibitem[{{Zink} {et~al.}(2020){Zink}, {Batygin}, \& {Adams}}]{zink20}
{Zink}, J.~K., {Batygin}, K., \& {Adams}, F.~C. 2020, Astronomical Journal,
  160, 232, \dodoi{10.3847/1538-3881/abb8de}

\end{thebibliography}
\bibliographystyle{aasjournal}

\end{document}